\documentstyle[preprint,aps]{revtex}
\tighten

\begin{document}

\title{Nucleon described by the chiral soliton \\
in the chiral quark soliton model}

\author{Teruaki Watabe
\footnote{E-mail address : watabe@hadron.tp2.ruhr-uni-bochum.de}
and Klaus Goeke}

\address{Institute for Theoretical Physics II, Ruhr-University Bochum, \\
D-44780 Bochum, Germany}
\maketitle

\begin{abstract}
We give a survey of recent development and applications of the chiral quark 
soliton model (also called the Nambu--Jona-Lasinio soliton model) with $N_f$=2 
and $N_f$=3 quark flavors for the structure of baryons.
The model is an effective chiral quark model obtained from the instanton 
liquid model of the quantum chromodynamics.
Mesons appear as quark-antiquark excitations and baryons arise as 
non-topological solitons with three valence quarks and a polarized Dirac sea.
In this model, a wide variety of observables of baryons is considered.
\end{abstract}

\section{Introdunction}

For many years effective chiral models have been used to calculate properties 
of baryons.
We have in mind theories like the Skyrme model, the Nambu--Jona-Lasinio and 
chiral quark soliton approach, chiral and cloudy bag models, etc.
In this paper, we briefly review recent development and applications of the 
chiral quark soliton model (one of the chiral models) with $N_f$=2 and $N_f$=3 
quark flavors for the structure of baryons.\footnote{Ref.~\cite{Review} gives 
a full detail of calculations.}

It has been believed that the quantum chromodynamics (QCD) describes the 
interaction between quarks.
However it is quite difficult to describe nucleons and baryons directly from
QCD.
The difficulty comes from the behavior of gluons in a low energy region.
Diakonov and Petrov have proposed the instanton liquid picture of the QCD 
vacuum.~\cite{DP86}
Through that picture the dynamical chiral symmetry breaking is induced.
Finally we have a quark-meson effective theory given by the Lagrangian;
\begin{eqnarray}
{\cal L} = \bar{\psi} ( i\gamma^\mu\partial_\mu -\hat{m} 
-Me^{i\gamma^5\lambda_a\phi^a/f} ) \psi
\ .
\label{eq:Lagrangian}
\end{eqnarray}
The chiral quark soliton ($\chi$QS) model is based on the Lagrangian 
(\ref{eq:Lagrangian}).
Using the saddle-point approximation of the meson field, baryons appear as a 
non-topological solitons.~\cite{DPP88}

The parameters, cut-off parameter of quark loop and current quark masses, are 
fixed to reproduce properties of mesons in the vacuum.
The coupling strength between quark and meson denoted by the $M$ is fixed to 
reproduce the $N-\Delta$ mass splitting and the fixed range is 
$M=400-450$~MeV.~\cite{Review}

\section{Nucleon Form Factors}

Nucleon form factors are given by the baryon matrix elements of quark bilinear 
operators;
\begin{eqnarray}
\left\langle P^\prime \left| V^\mu_{\cal Q}(0) \right| P \right\rangle = 
\left\langle P^\prime \left| \bar{\psi}(0) \Gamma^\mu \lambda_{\cal Q} \psi(0) 
\right| P \right\rangle
\ .
\label{eq:bilinear}
\end{eqnarray}
In the $\chi$QS model the matrix elements (\ref{eq:bilinear}) consist of 
{\it valence} and {\it sea} parts.
The sea part can be recognized as a contribution from mesonic clouds through 
the derivative expansion of the meson filed in a limit of large coupling 
strength $M$.~\cite{WKG96}

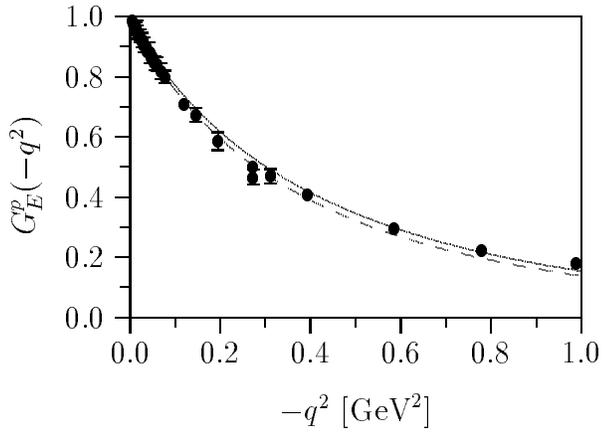
\begin{figure}[hp]
\caption{The experimental data are from ref.~{\protect\cite{Hoh}}.}
\label{fig:elp}
\end{figure}

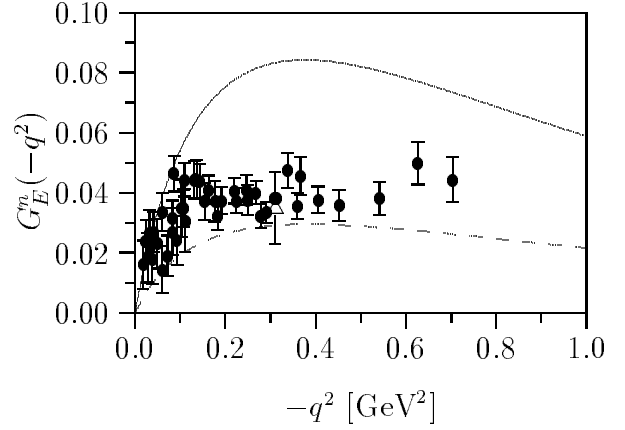
\begin{figure}[hp]
\caption{The experimental data are from ref.~{\protect\cite{Pla}} denoted by 
solid circles and ref.~{\protect\cite{Mey}} denoted by an open triangle.}
\label{fig:eln}
\end{figure}

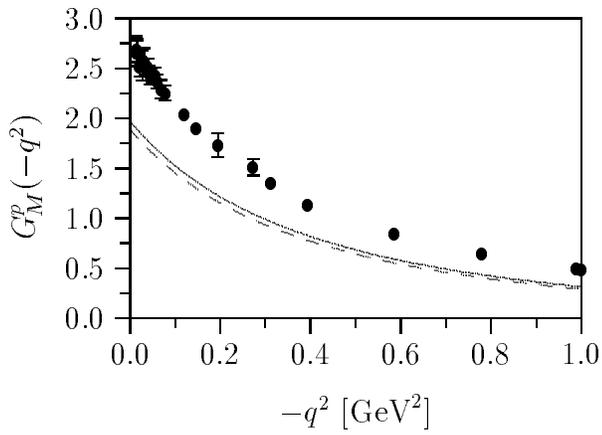
\begin{figure}[hp]
\caption{The experimental data are from ref.~{\protect\cite{Hoh}}.}
\label{fig:mgp}
\end{figure}

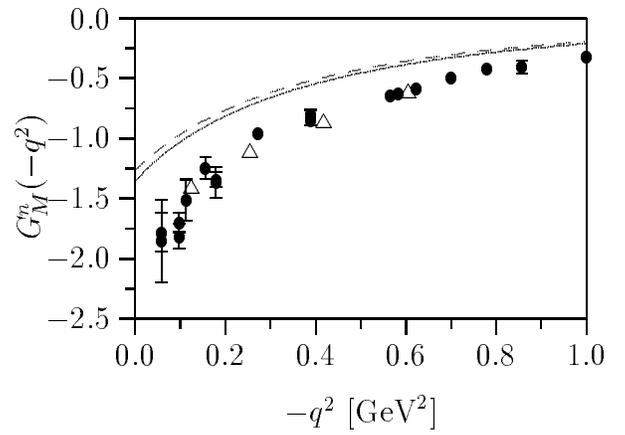
\begin{figure}[hp]
\caption{The experimental data are from ref.~{\protect\cite{Hoh}} denoted by 
solid circles and ref.~{\protect\cite{Bru}} denoted by open triangles.}
\label{fig:mgn}
\end{figure}

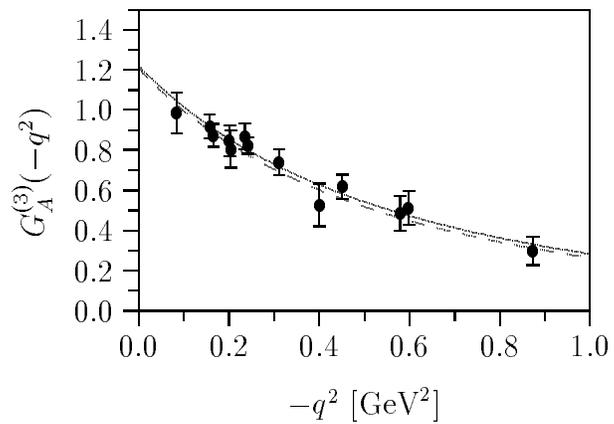
\begin{figure}[hp]
\caption{The experimental data are from ref.~{\protect\cite{Bak,Kit}}.}
\label{fig:ax}
\end{figure}

We show results of form factors, which are calculated with $M=420$~MeV, in 
Figure~\ref{fig:elp}-\ref{fig:ax}.
The solid lines correspond to results of SU(2)$_f$ calculations.~\cite{Review}
On the other hand, the dashed lines are results in SU(3)$_f$.~\cite{WG97}
One could find that the results of SU(3)$_f$ calculations in 
Figure~\ref{fig:elp}-\ref{fig:ax} are different from the results in 
ref.~\cite{Review}.
The reason is the following.
The SU(3)$_f$ calculation is based on the embedding ansatz of SU(2)$_f$ soliton
into SU(3)$_f$.
To quantize the soliton, we rotate it in the SU(3) flavor space.
Because the soliton is not a symmetric object in the SU(3) flavor space, there 
are off-diagonal elements of moment of inertia.
In ref.~\cite{Review} the authors symmetrize the matrix of moment of inertia 
to quantize the soliton.
However it causes to a problem of the electric charges of baryons.
Usually contributions to the baryon matrix elements are given by 
multiplications of dynamical- and collective-parts.
In a calculation of the electric charge of baryon, dynamical parts should be 
the same as the moments of inertia and they are exactly normalized by the 
moments of inertia.
But due to the artificial symmetrization of the matrix of moment of inertia, 
the normalization is not done perfectly.
Some non-integer terms remain.
The authors of ref.~\cite{Review} neglect those non-integer terms by hand, 
while they keep full terms in the other calculations like magnetic moments and 
axial coupling constants.
Hence the SU(3)$_f$ calculations in ref.~\cite{Review} involve some 
inconsistency.
To avoid that inconsistency we diagonalize the rotation of the SU(3)$_f$ 
soliton.~\cite{WG97}
By the diagonalization of the rotation, we have obtained the results shown in 
Figure~\ref{fig:elp}-\ref{fig:ax} and they are almost the same as the SU(2)$_f$
results except the neutron electric form factor.

\section{Discussion}

In Figure~\ref{fig:elp}-\ref{fig:eln} the electric form factors of proton and
neutron are given.
In the $\chi$QS model the proton electric form factor is dominated by the 
valence contribution, while the neutron is by the sea contribution which is 
interpreted as a mesonic cloud contribution.
The difference between SU(2)$_f$ and SU(3)$_f$ exists only on the neutron 
electric form factor and it can be explained by the following two reasons.

1) It could be due to a systematic error of the model.
Because the neutron electric form factor is quite tiny compared with proton's, 
it is probably sensitive to a systematic error of the model.

2) It could be due to a kaonic excitation of the vacuum.
In SU(2)$_f$ the vacuum excitation in neutron is via a $\pi^-$ channel.
On the other hand, in SU(3)$_f$, it is via not only a $\pi^-$ channel but also 
a $K^+$ channel.
The $K^+$ excitation suppresses the $\pi^-$ excitation.
Because of that, the SU(3)$_f$ result is below the result in SU(2)$_f$.

Within the present formalism it is not possible to come to a conclusion.
We should perform an accurate calculation which allows us to control 
properties of the kaon, for instance the kaon mass.
One possible method is given by the bound-state approach proposed by Callan and
Klebanov in the Skyrme model.\footnote{It is in progress by T. Watabe, C. Vogt 
and K. Goeke.}~\cite{CK}

\section{Summary}

We have briefly shown recent development and applications of the chiral quark 
soliton model with $N_f$=2 and $N_f$=3 quark flavors for the structure of 
baryons.
The results do not have a remarkable difference between SU(2)$_f$ and SU(3)$_f$
except the neutron electric properties.
The difference in the neutron electric properties could be interpreted as an 
effect of the vacuum excitation in the $K^+$ channel.
To come to a conclusion we need a further investigation.

\end{document}